\documentclass{sigcomm-alternate}
\usepackage{graphicx}
\usepackage{url}
\usepackage{booktabs}
\usepackage{subfigure}
\usepackage{color}

\begin{document}

\title{``Resource Pooling'' for Wireless Networks: \\Solutions for the Developing World}
%\title{Cognitive Radios for Development (C4D)}
\author{\large Junaid Qadir$^{1}$, Arjuna Sathiaseelan$^{2}$, Liang Wang$^{2}$, Jon Crowcroft$^{2}$\\
%Marco Zennaro$^{3}$, Adam Wolisz$^{4}$ \\
\normalsize $^{1}$Information Technology University (ITU)-Punjab, Lahore, Pakistan\\
\normalsize $^{2}$Computer Laboratory, University of Cambridge, United Kingdom\\
%\normalsize $^{3}$The Abdus Salam International Centre for Theoretical Physics (ICTP), Italy\\
%\normalsize $^{4}$Technische Universit{\"a}t Berlin, Germany\\
\normalsize Email: junaid.qadir@itu.edu.pk; arjuna.sathiaseelan@cl.cam.ac.uk; liang.wang@cl.cam.ac.uk; jon.crowcroft@cl.cam.ac.uk\\}
%mzennaro@ictp.it; awo@ieee.org\\}
\maketitle

\begin{abstract}

We live in a world in which there is a great disparity between the lives of the rich and the poor. Technology offers great promise in bridging this gap. In particular, wireless technology unfetters developing communities from the constraints of infrastructure providing a great opportunity to leapfrog years of neglect and technological waywardness. In this paper, we highlight the role of resource pooling for wireless networks in the developing world. Resource pooling involves: (i) abstracting a collection of networked resources to behave like a single unified resource pool and (ii) developing mechanisms for shifting load between the various parts of the unified resource pool. The popularity of resource pooling stems from its ability to provide resilience, high utilization,  and flexibility at an acceptable cost. We show that ``resource pooling'', which is very popular in its various manifestations, is the key unifying principle underlying a diverse number of successful wireless technologies (such as white space networking, community networks, etc.). We discuss various applications of resource pooled wireless technologies and provide a discussion on open issues.

\end{abstract}

\section{Introduction}

%
%\begin{quotation}
%``\textit{We are realizing that access to the Internet is not just a luxury. The gap between those who are connected and those who are not is so large that if you disconnect someone's house it is a little bit like imprisoning them.}''---Tim Berners Lee.
%\end{quotation}
%

%The United Nations Millennium Declaration in September 2000 vowed to spare no effort to free our fellow men, women, and children from the abject dehumanizing conditions of extreme poverty currently faced by more than one billion people. Despite a lot of effort, poverty is still an unresolved problem and approximately 13 percent of the world's population are living in abject poverty (World Bank estimates, 2012). 

It has been shown that the Internet has the potential to foster development and growth, but this potential is thwarted by the fact that billions of people are unable to access the Internet \cite{thanki2012economic}. People not having Internet access---nearly half of whom live below the poverty line in their respective countries---are more likely to be uneducated, poor, and residents of undeveloped rural areas. These availability and affordability gaps are disproportionately impacting people in Africa, Asia, and Latin America and contributing to a ``digital divide'' between humanity. Bringing the Internet to the remaining billions of people will democratize knowledge, open up new opportunities, and undoubtedly open up avenues for sustained development. The fact that Internet access can play a large role in facilitating development motivates the vision of \textit{Global Access to the Internet for All} (GAIA), currently being formally pursued in the Internet Research Task Force (IRTF). 

%In worldwide terms, there are more people who cannot access the Internet than there are who can (the majority of the world's popultion (around 60\%) are currently of the world's population is offline \cite{thanki2012economic}).  

%Technology can be a potent leveling tool in the hands of the underdeveloped that has the empowering ability to bridge the gap between the quality of lives of the rich and the poor. 

%In particular, Internet access is a key indicator of the potential of economic progress. Internet's impact is imprinted on all spheres of human life---personal, societal, political, economical,  and educational---in both developing and developed countries. 

%\footnote{For more than 60\% of the world's population (and up to 90\% of the population of Africa), the price of the Internet is unaffordable.}\cite{thanki2012economic}

While traditionally fixed wired broadband connections have been used to provide Internet to the majority of online population in the developed countries, this approach is not attractive for developing countries due to the high cost of wiring and infrastructure. The emergence of \textit{mobile phones} and \textit{wireless technology} allows the people in developing regions an opportunity to bypass the need of fixed-line broadband and personal computers and leapfrog towards an mobile broadband offering \cite{gao2004wireless}. 

While mobile technology has been one of the biggest success stories of engineering, cellular technology---which is employed by (for-profit) telecom after deploying terrestrial infrastructure such as base stations---is not well suited for provisioning global Internet/ connectivity. Since sufficient user density is needed to justify the cost of installing expensive infrastructure, mobile connectivity is mostly restricted to urban areas. In addition, traditional cellular companies are also hard pressed to scale to the exponentially increasing performance demands (due to exponential rise in the number of mobile users and their bandwidth demands).

While the use of wireless technology in the developing countries has been proposed in a number of papers \cite{Gunasekaran200723} \cite{subramanian2006rethinking}, traditional wireless networks suffer from a number of performance, scalability, and economic viability problems. In this paper, we argue that \textit{resource pooling based wireless} technologies, through its various manifestations (elaborated in Section \ref{sec:manifestations}), can provide a solution to the problems associated with traditional wireless. The main idea underlying ``resource pooling'' is to develop an abstraction of a singled pooled resource that behaves as a single queue even when the resources constitute a disparate collection. Resource pooling techniques characteristically operate in a bottom-up fashion, which has been shown to be an effective operating principle for  developmental work \cite{easterly2008institutions}. Our work is the first---to the best of our knowledge---that highlights resource pooling as a general underlying architectural principle that can be used to improve the appropriateness of wireless technology for developing regions while also improving its efficiency and performance.

The organization of this paper follows. We describe the various manifestations of resource pooling in Section \ref{sec:manifestations}. We describe how the resource pooling principle underpins diverse wireless technologies in Section \ref{sec:rpWireless}. We discuss the various open issues related to resource pooled wireless technologies in Section \ref{sec:discussion}. We conclude this paper in Section \ref{sec:conclusion}.

% with a summary of this paper's main contributions and insights.

\section{Resource Pooling Manifestations} 
\label{sec:manifestations}

%In this paper, we propose the idea that resource pooling is an important unifying concept when we consider wireless-based technological solutions for developing areas. Resource pooling is naturally suited to the developing world, where maintaining dedicated IT infrastructure is especially cost prohibitive for small-scale entrepreneurs, business owners, and non-profits. The general idea of resource pooling encompasses modern trends of dynamic spectrum access networks (that utilize the so called spectrum ``white space''), community wireless networks, multihoming with heterogeneous wireless technologies, and mobile cloud computing. 

%On the basis of these attributes, we can categorize wireless resource pooling techniques into techniques that i) use aggregation/ concurrency/ bonding; ii) use centralization of resources; iii) dynamically allocate resources; iv) load balance in time or space to control congestion; v) support resource sharing/ multi-tenancy/ virtualization; and are vi) based on community sharing/ crowdsourcing. 

%Resource pooling has influenced and shaped a number of successful Internet technologies such as packet switching, peer-to-peer networking (P2P), content-delivery networks (CDNs), and cloud computing. 

Although resource pooling has many manifestations in wireless technology, these manifestations can be categorized according to seven principal signature attributes. A description of this classification---which is not mutually exclusive, i.e., a technology can embed multiple resource pooling manifestations---follows and is summarized in Table \ref{tab:MP_advantages}. 

\subsection{Techniques Based On Aggregation/ Concurrency/ Bonding}

A common type of a resource pooling technique depends on bonding or aggregating multiple resource together to act like one bigger more powerful resource.   Such a technique also can utilize the aggregated resources concurrently. Examples of this type include the use of multiple paths to communicate with a destination; the use of multi-input multi-output (MIMO) techniques; and bonding multiple channels together, as in Wi-Fi bonding, and 3G bonding.

\begin{table}[!ht]
\centering
\scriptsize
%\tiny
\caption{Wireless Resource Pooling Technique Types and Examples}
\label{tab:MP_advantages}
\begin{tabular}{p{8.3cm}}
\hline
\\

\textbf{1. Techniques that use \underline{aggregation/ concurrency/ bonding}}:\\
Multipathing (over multiple wireless interfaces: e.g., 3G/ Wi-Fi)\\ 
%Joint multiuser beamforming (JMB) \cite{rahul2012jmb} \\
MIMO; Wi-Fi bonding \cite{bukhari2016survey}; 3G bonding \cite{sathiaseelan2013internet};\\ 
%FatVAP \cite{kandula2008fatvap}\\
%Non-contiguous Orthogonal Freq. Division Multiplexing (NC-OFDM)  \cite{rajbanshi2006efficient}\\

\\
\textbf{2. Techniques that use \underline{centralization} of resources}:\\
Radio Resource Management (RRM)\\
Cloud radio access networks (C-RAN) \cite{checko2014cloud}\\
Mobile Virtual Network Operators (MVNOs)\\
Wireless spectrum resource pool (WSRP) \cite{yang2013openran}\\

\\
\textbf{3. Techniques that use \underline{dynamic allocation of resources}}\\
% (instead of static allocation)}:\\
Dynamic Spectrum Access (DSA); White Space Networking\\
Less than Best Effort (LBE) Service \cite{shalunov2010low}\\
%QBone Scavenger Service

\\
\textbf{4. Techniques for \underline{load balancing} to control congestion}:\\
\\
\textit{\textbf{Load Balancing Over Paths/ Channels/ Peers}}:\\
Multihoming; Multichannel wireless\\
Multipath routing and transport\\

\\
\textit{\textbf{Load Balancing Over Time}}:\\
Low Extra Delay Background Transport (LEDBAT) \cite{shalunov2010low}\\
Delay Tolerant Networking \cite{fall2003delay}; Pocket Switched Networks \cite{chaintreau2005pocket}\\

\\
\textbf{5. Techniques for \underline{Resource sharing/ multi-tenancy/ virt.}}
\\
Multiplexing, Packet Switching, DSA\\ 
Virtualization, Cloud Computing, MVNOs\\
%Spectrum Without Bounds, Networks Without Borders \cite{doyle2014spectrum}\\
Virtualization (of BTS; VMs, VNs) \cite{doyle2014spectrum}\\
%Multi tenant mobile networks \cite{qadir2014building};\\
Virtualization of 4G/ 5G RAN; Wi-Fi APs \& NICs \cite{qadir2014building};
%Wireless Virtualization \cite{doyle2014spectrum}\\

\\
\textbf{6. Techniques based on \underline{community sharing/ crowdsourcing}}:\\
Wi-Fi offloading \cite{lee2010mobile}\\
Community Wireless Networks: LCD-Net \cite{sathiaseelan2013lcd}; Public Access Wi-Fi Service (PAWS) \cite{sathiaseelan2014feasibility};\\% OpenWifi \cite{hu2007openwifi};\\
Community GSM Networks \cite{hasan2014gsm};\\
Fog Computing \cite{bonomi2012fog}\\

%\\
%\textbf{7. Techniques that use \underline{virtualization}}:\\
\\
\textbf{7. Techniques for \underline{versatile resource usage}}:\\

Software Defined Radio (SDR); Network Coding;\\
TCP Hierarchical Acknowlegement (HACK) \cite{salameh2014hack}\\

\\
\hline
\end{tabular}
\end{table}

\subsection{Techniques Using Centralized Resources}

There has been a perennial tension between the distributed and the centralized processing and management paradigms. While distributed processing allows for more robust localized implementations, it is also true that centralized schemes guarantee optimal resource pooling (albeit at the cost of additional communication overhead and lower fault tolerance). It has been shown by Tsitsiklis et al. \cite{tsitsiklis2012power} that resource pooling benefit can be exploited even with a  small degree of centralization leading to exponential performance improvements in steady-state scaling of system delay for sufficiently large systems. 

\subsection{Techniques For Dynamic Allocation of Resources}

%To put things in perspective, circuit switching techniques use fixed multiplexing schemes in which dedicated and isolated non-pooled circuits are utilized.  
 
Decades of experience with Internet has reinforced a general rule of thumb: it is nearly always more preferable to dynamically allocate resources (a concept that complements ``resource pooling'') over static resource allocations. Dynamic resource allocation can allow for more efficiency and flexibility in situations where the demand on scarce resources is not predictable. The \textit{packet switching} technology, which acts as a foundation of the Internet, is essentially a resource pooling technique. Packet switching uses \textit{statistical multiplexing} to achieve resource pooling through dynamic resource allocation by allowing a burst of traffic on a single circuit to use spare capacity on other circuits. More specific for wireless networks, the ``\textit{dynamic spectrum access}'' (DSA) technique (elaborated in Section \ref{dsa}) is another dynamic resource allocation technique. The DSA model tolerates an unlicensed, or the secondary user (SU), in licensed spectrum subject to the provision that the licensed users, or the primary users (PUs), are not interfered with. Another example of dynamic allocation of wireless resource pool is the use of \textit{unlicensed spectrum} to unlock spectrum (which in licensed spectrum would be dedicated for particular licensed applications/ users). The use of unlicensed spectrum has buoyed an increased interest in using Wi-Fi, and similar unlicensed technologies, for bringing Internet access to the developing world (and hundreds of communities are already operating municipal broadband wireless network that provides broadband access to the Internet \cite{mandviwalla2008municipal}). 

%Referring to our taxonomy (Table 1), resource pooling manifests itself with unlicensed spectrum in the form of aggregating resources (and not balkanizing them by dedicating them to licensed users) and dynamic resource allocation.  

%The traditional command-and-control based spectrum allocation approach---that statically allocated portions of the spectrum to different licensed users/ applications---has been shown to be grossly inefficient. The traditional approach leads artificially to spectrum scarcity---even though the spectrum is mostly available, it is off limits for new applications due to the static licensing that precludes non-licensed users from accessing the spectrum. 

%The use of unlicensed frequency is an important factor leading to the popularity of Wi-Fi. 

%Indeed, IEEE 802.11, or Wi-Fi, is already carrying the bulk of world's data traffic including approximately 57\% of total traffic generated by PCs and laptops (which is more than the combined total of the shares of cellular and wired connections) and a bigger share of data generated by smartphones/ tablets (69\% of total traffic) \cite{thanki2012economic}. 

\subsection{Techniques Relying On Load Balancing}

Load balancing and resource pooling are also closely related. The efficiency and the power of the pooled resource depend on how the load is balanced over the individual resources. Load balancing can be temporal (i.e., the load is balanced between peak and non-peak hours), spatial (i.e., the load is balanced between different locations), or it can be over multiple resources (e.g., the load is balanced over multiple links, as in multihoming, or over multiple paths as is done in multipath transport protocols). This load balancing can be over paths/ channels/ peers or over time (with more illustrative examples provided in Table 1).

\subsection{Techniques For Resource Sharing/ Multi-Tenancy}

One of the main purpose of the original Internet was to lower the cost of computing by networking the time-sharing systems that existed at many universities. The same aim of resource sharing and multi-tenancy is driving the modern trends of cloud computing and virtualization. Cloud computing relies on pooled resources gathered at centralized datacenters to support the computing requirements of multiple tenants through virtualization. The multi-tenant model of cloud computing is closely related to resource pooling: in a cloud, each tenant is assigned, and reassigned as needed, all the physical and virtual resources it needs per its demand. 

\subsection{Techniques Based on Crowdsharing/ Community Sharing}

A number of diverse wireless technologies---such as mobile crowdsensing \cite{lane2010survey}---have been developed that can effectively pool the increasing capabilities of mobile devices to provide business, healthcare, environmental monitoring, transportation, and social networking services. Other wireless resource pooling techniques include crowdsourced efforts such as Wi-Fi offloading aimed at improving the scalability and performance of mobile networks \cite{lee2010mobile}. Community networking techniques are discussed in more detail later in Section \ref{sec:communityNet}.

%, and cooperative techniques for low power low latency cellular connectivity \cite{nicutar2014using}. 

\subsection{Techniques For Versatile Resource Usage}

The Latin word \textit{versatilis} connotes turning, or having the capable of turning to varied subjects or tasks. One aim of resource pooling is to versatilely apply resources in diverse on-demand, and not predefined, configurations. Versatile resource usage pools the same resource for many different purpose. A good example of a versatile resource pooling technique is software defined radio (SDR). SDRs by their versatile nature are radio chameleons that can use software programmability to pool various wireless networking technologies onto the same radio resource---e.g., SDR may run a telephony protocol (e.g. CDMA) at one time and a totally distinct data communication protocol (e.g. Wi-Fi) at another. Other examples of versatile resource pooling include network coding (which pools multiple messages into a smaller number of coded messages by more versatilely defining the routing forwarding action) and cross-layer optimization that pools multiple acknowledgments (ACKs) into a single ACK making ACKs more versatile. 

%In this fashion, SDR uses the pooled hardware resource to generate different kinds of radios. 

\section{Resource Pooled Wireless \\(Wireless 2.0)} 
\label{sec:rpWireless}

%Resource pooling technologies have many different manifestations. 

We will next cover a number of important wireless resource pooling 
techniques that will likely form the basis of ``Wireless 2.0'', the next generation of wireless technology that will have better efficiency and also more advanced functionality. In addition to their role in advancing the state of the art, these technologies are also amenable to deployments in developing environments (due to their bottom-up resource-pooled nature). 

\subsection{Dynamic Spectrum Access}
\label{dsa}

Dynamic spectrum access (DSA) refers to the set of technologies that can enable radios to opportunistically access radio spectrum (that may be licensed but is currently unused) to effectively create a ``secondary network'' on top of an existing ``primary network''. In networks that use DSA, also known as ``white space networks'', nodes can increase the efficiency of spectrum utilization through ``spectrum pooling'' \cite{weiss2004spectrum} and thereby avoid the problem of ``artificial spectrum scarcity'' inherent to the command-and-control licensing approach to spectrum that can deny consumers and their devices wireless bandwidth even when no one else is using it. Just like packet switching improved upon circuit switching, DSA can improve the command-and-control spectrum licensing approach by sharing the radio spectrum through statistical multiplexing. However, the increased flexibility due to DSA comes with many technological challenges with the primary challenge being ensuring protection to ``primary users'' (PUs). 

%With the emergence of IoT technology (which promises to connect potentially billions of interconnected wireless devices and thereby create new disruptive applications), there is a strong motivation to engage in research on DSA that will facilitate the wireless internetworking required without exacerbating the spectrum crunch.

%\vspace{1mm}
%\subsubsection{TV whitespaces (TVWS)}

There are two important types of white space networks: TV white space (TVWS) networks and GSM white space networks. \textit{TVWS} is composed of the VHF and UHF band spectrum that was traditionally earmarked for TV transmissions but was unused. There is an increasing trend globally of opening these TV white spaces (which have very desirable transmission properties) for broadband access. This trend is being accelerated by the switch from analogue to digital television (which takes less spectrum). Broadly speaking, there are four key reasons that ICT for development (ICTD) community is excited about TVWS in developing countries \cite{Zennaro2013tvws}: i) low risk regulation; ii) suitability for rural environments; iii) ample availability of TV spectrum in rural/ underdeveloped areas; and iv) opportunities for entrepreneurship (through which the wireless service market has been democratized).  

With the emergence of low-cost software defined radios (SDRs), along with the availability of open-source software such as OpenBTS\footnote{\url{http://openbts.org}}, a growing recent trend is to utilize \textit{community cellular networking} to support rural area cellular networking using \textit{GSM white spaces}. While building and operating top-down cellular technology, as operated by traditional mobile companies, is very expensive (especially for rural settings),  a rural community can support itself with community networking at a very reasonable cost. The trend of using local community networks with unlicensed spectrum promises to ameliorate the issue of digital exclusion of rural communities \cite{heimerl2013local}.

\subsection{Community Networking}
\label{sec:communityNet}

It has been shown by researchers that the adoption of an indigenous bottom-up approach is the key to sustainable development \cite{easterly2008institutions}. Community networks follow the bottom-up philosophy and are ``run by users, for the users''. Community-based resource pooling can be in the form of an extension of a user's paid service (as in BT FON\footnote{\url{http://www.btfon.com}}, where home users publicly share their home broadband connections in return of credits that may be used to access other user's APs) or in the form of a cooperative wireless mesh network (e.g., Guifi\footnote{\url{http://guifi.net}}, which is an open self-organizing community network that uses unlicensed wireless links and open optical fiber links. The applications of community networking extend to much more than Internet access: e.g., a number of community communication scenarios can be entertained such as local intranets and machine-to-machine networks that provide localized services, such as neighborhood groups and surveillance, high-speed P2P networking, high-speed video conferencing, local TV/radio broadcasting stations. While community networking also has application in urban settings, the biggest draw for community networking is in rural settings (where cellular networks and satellite networks are expensive to operate). If suitably incentivized, crowdsourced resource pooling has the potential of mitigating the digital exclusion of rural populations by breaking the impasse emerging from the mismatch between the economic (profit-seeking) interests of commercial service providers and the egalitarian goal of GAIA.

\subsection{Multihoming with Heterogeneous Networks}

In the not too distant past, multihoming was relevant only for critical servers, but with billions of mobile now multihomed with heterogeneous access technologies, multihoming has now become mainstream. There is a lot of ongoing research on supporting heterogeneous multihoming with work aiming to support (i) seamless handover between mobile cellular connections and Wi-Fi \cite{paasch2012exploring}, (ii) supporting scalable wireless networking through Wi-Fi offloading and 3G onloading, and (iii) concurrent usage, or bonding, of data and telecom connections for higher performance \cite{deng2014wifi}. A modern multihomed wireless device can exploit its interfaces (such as 3G/ Wi-Fi) to improve its reliability, throughput, latency, and fault tolerance. In recent times, there have been multipath transport level initiatives (such as MP-TCP) that seek to support such multihoming by allowing a client to establish multiple connections to the same destination host over different network adapters. Siri on Apple's iOS7 uses MP-TCP to improve its usability by making its service more resilient to network failure. This is managed by opening two connections: the primary TCP connection is created over Wi-Fi while the backup connection---which is used when Wi-Fi becomes unresponsive---works using the cellular connection. 

%With future mobile networks expected to carry ever-increasing amounts of data, mobile network providers are increasingly turning to Wi-Fi offloading as a technique for relieving the burden on cellular networks by offloading the traffic through local Wi-Fi connections. Apart from Wi-Fi offloading, there have also recent efforts that have explored 3G onloading to complement home broadband networks with 3G data connections to boost throughput and performance.

%{\color{red}Exploring mobile/WiFi handover with multipath TCP \cite{paasch2012exploring}
%
%WiFi, LTE, or Both?: Measuring Multi-Homed Wireless Internet Performance .}

\subsection{Mobile Cloud Computing}

The mobile cloud computing trend seeks to exploit the great promise of cloud computing and the ubiquity of mobile smartphones. Apart from the benefits of scalability, performance, and efficiency, using cloud computing in developing countries is attractive for another important reason. Users in developing countries can combine cheap battery-powered smartphones with cloud servers to circumvent the power problems due to unreliable intermittently available electrical grid. In addition, the lack of existing infrastructure implies that users in developing countries can directly adopt advanced cloud/ wireless technology without being encumbered by infrastructure conversion costs \cite{greengard2010cloud}. Assuming the availability of a decent Internet connection, the ``software as a service'' model is also attractive in the developing world as a substitute for in-house services like email, file sharing, etc. In addition, a world of opportunities in fields as diverse as mobile health,  education, banking, agricultural support exist (through mobile paradigms dubbed as \textit{m-Health}, \textit{m-Education}, \textit{m-Banking}, and \textit{m-Agriculture}) \cite{greengard2010cloud}. In related work, there has been a lot of recent efforts to bring the benefits of cloud computing to cellular and wireless networking. In particular the ``\textit{network without borders}'' (NWoB) architecture presented in  \cite{doyle2014spectrum} envisions a pool of resources (comprising aggregated spectrum and infrastructure resources) that are used and repurposed as needed to support the instantiation and orchestration of multiple virtual wireless networks. 

%Such a network architecture can meet the challenges of meeting the ever-increasing demand of bandwidth and spectrum scarcity in a cost-effective fashion.

%Mobile cloud computing is also well suited to support public health services. As an example, the open source mobile health (m-health) technology suite MOTECH suite which works with cheap Java based phones can be used to connect health workers with patients. 

%In the EU funded Trilogy 2 project, the metaphor of a ``\textit{liquid system}'' puts forward the vision of a system that allows an application to use, and scale up and down on demand, resources from a common resource pool created from bandwidth, storage, and processing contributions by network operators, datacenter operators, and end systems \cite{trilogy2}. 

%The risk of miscalculating resource need shifts to cloud providers, a significant advantage if the resource demand varies a lot or is just hard to predict. 

\subsection{Virtualized Wireless Networks}

Virtualization has transformed the efficiency of the datacenter environment by enabling cloud computing and promises to bring a similar revolution to wireless networking. The trend of virtualized wireless networks (VWNs) is fast emerging since it brings the performance, convenience, and cost-efficiency of virtualization to wireless networking \cite{qadir2014building}. VWNs can facilitate the convergence and interworking of different wireless technologies and enable efficient resource sharing---e.g., VWNs can be used to support multiple virtual network operators (VNOs), using which mobile VNOs can share the cellular infrastructure and thereby lower its capital and operational expenditures (CAPEX and OPEX). In recent times, there have been efforts of virtualizing WLANs, cellular networks, SDN-based wireless networks, as well as cloud-based VWNs and cognitive-radio-based VWNs \cite{qadir2014building}. There is also the emerging trend of network functions virtualization (NFV), which  promises to disrupt the telecom industry by decoupling the functionality of telecom devices from its physical instantiation in dedicated telecom hardware. 

%and utilize the increasing computing/ storage resources available at modern smartphones located near the edge. , there is an interest in using localized computing resources to avoid/mitigate the bandwidth limitations of wireless networks. Fog computing is an effort in the general trend of pushing computing/ storage/ caching closer towards the edge \cite{bonomi2012fog}. 

%With the emergence of the big data paradigm due to the proliferation of sensing devices and embedded computers, wireless networks (such as cellular networks)---especially in developing countries---are simply not fast enough to transmit data from generating devices to cloud data centers at the rate of generation of data. 

\subsection{Edge Computing/ Fog Computing}

The two main driving forces of edge/ fog computing are: i) the undesirability of moving large amounts of data to the cloud, especially through low-bandwidth wireless links; and ii) the torrential amounts of data that will increasingly characterize internet in the era of IoT. With edge computing/ fog computing, the resource pool at the edge (comprising of the computing, storage, and communication resources of user's smartphones/ tablets) is utilized instead of the central resource pool that may be hosted in a cloud datacenter \cite{bonomi2012fog}. By exploiting the edge resources and avoiding the reliance on the cloud, we can reap the benefits of better responsiveness and a wider service coverage. The recent trend of device-to-device (D2D) communication---defined in the context of cellular mobile networking as the direct communication between two mobile users (also called UEs or User Equipment) without traversing the base station or the core cellular network---works in a similar vein. D2D communication technology can be a valuable fallback for public safety communications in the case of disasters and emergencies in which some  (or all) of the cellular networking infrastructure becomes fails or becomes unavailable. The use of D2D communication in 3G, 4G, and 5G technology is currently being actively pursued.  

%The 3GPP body is particularly interested in using D2D in LTE for public safety networks \cite{lin2014overview}. 

%This trend of storing and processing the data on the devices locally is a good fit for the upcoming IoT era. Such a network architecture is sometimes termed as fog computing (to emphasize the close proximity of computing to the devices in contrast with cloud computing which is comparatively further away). 

%In recent times, fog computing is going mainstream in wireless networking. Fog computing, or the reliance on edge devices, is also important in challenged networks such as those that characterize the developing world. 

\subsection{Delay-Tolerant Networking (DTN)}

Since it is anticipated that many alternative networking technologies will play a part in universal Internet access, the always-connected connectivity model of the Internet does not lend itself as an economically viable solution for GAIA. DTN is a temporal resource pooling technique that can be used to transport traffic over time even when end-to-end connectivity is intermittent. Originally developed as an architecture for the interplanetary Internet \cite{burleigh2003delay}, delay tolerant networks (DTNs) are now recognized as an alternative solution for building disruption tolerant solutions in systems where the communication links are intermittently available (e.g., as in LEO constellation system). DTN communication paradigm aims to provide a less-than-best-effort service and aggressively widen the possible connectivity options beyond those available currently.  

\subsection{Information-Centric Networking (ICN)}

The basic insights of ICN are (1) the Internet operation should be focused on the exchange of content (without being coupled to the location of the communicating end hosts), and (2) there should be universal caching embedded in the architecture. ICN proposes to replace the the send-receive paradigm of the current Internet (that couples the location/identifier of the communicating machines) will be replaced by a publish-subscribe paradigm (that provides an architectural spatial and time decoupling). ICN promises benefits such as lower response time (due to pervasive caching and nearest-replica routing) and better support for traffic engineering, content distribution, security, and mobility. ICNs can provide many of the benefits of CDNs (such as providing the requested data from close locations) without the drawbacks of CDNs (which are often implemented using proprietary technologies as overlays---and thus cannot utilize the underlying network's multicast and broadcast capabilities).  Similarly, ICN also leverages insights and technology from P2P networking (such as the use of distributed hash tables (DHTs)). 

%ICN is a promising GAIA architecture since it can reduce the access cost per bit through its emphasis on content caching both in the network and at the edges \cite{sathiaseelan2013information}. 

%While intermittent connectivity can result for various reasons, there is an underlying topological stability in the case of rural networks with the network dynamics occurring mostly due to link failure/congestion or unreliable power rather than node mobility. The network dynamics are also predictable especially in Approaches that use data ferrying (using buses or data mules). 

%DTNs are thus quite different from conventional networks and MANETs and require a distinctive design. A review of DTN routing protocols, and their design space, discussed in depth in \cite{ali2010routing}.

%A possible approach that seeks to redress this is to use opportunistic delay/disruption tolerant networking (DTN) communication paradigm to provide a less-than-best-effort service and aggressively widen the possible connectivity options beyond those envisioned currently.  

\section{Discussion}
\label{sec:discussion}

\subsection{Resource Pooling in Time and Space}

By employing resource pooling in time and space, we can unshackle the close coupling of the various disparate entities in the original Internet architecture---such as the close coupling of i) the host identifier and location in the case of IP address, and of ii) the content requested and the location of a server hosting this content---that hinders a flexible evolution of Internet's architecture to accommodate new applications such as video and mobility. In recent times, researchers have proposed many architectures that aim to redress such couplings. The DTN architecture aims to provide a decoupling in time, while the ICN architecture aims to provide a decoupling in space. Such decoupling is especially relevant in the context of developing countries and remote areas where end-to-end communication may not be possible and localized opportunistic communication needs to be used more leveraging ideas that follow from spatial and temporal decoupling of communicating nodes. By disrupting the traditional assumption of always-on Internet connectivity, nodes in such challenged networks can communicate using e.g. time-shifted access (in which delay-tolerant asynchronous bulk data is transferred at off-peak hours). In such scenarios, batched transfer (in contrast to always-on end-to-end transfer) can also lead to energy savings since a receiver does not have to continuously check for and process arriving packets. In recent times, researchers have focused their attention on exploiting the resouce pooling in time and space through the DTN/ ICN hybrid for enabling global access to the Internet for all (GAIA) \cite{trossen2016towards}.

\subsection{Where, When, and How to Resource Pool?}

\vspace{1mm}
\subsubsection{Centralized or Distributed Management}

%the spectrum used by IEEE 802.11 can be construed as resource pooled since the channels are not dedicated to particular users/ applications

While the resource pool incorporates centralization and aggregation of resources, the management of the resource pool can be done in a centralized or distributed fashion. Radio resource management (RRM)---which performs system-level control of radio resources and parameters in technologies such as cellular, wireless, and broadcasting networks---can be performed in both centralized and distributed fashion (although central control is more common). The modern software-defined networking (SDN) trend however is based on the centralization of the network control functionality in a centralized controller. 
 
%To illustrate this consider IEEE 802.11 Wi-Fi which uses 3 orthogonal, and 11 partially overlapping, channels in the 2.4 GHz ISM band. There are two ways of managing the spectrum resource pool: either through the centralized \textit{point coordination function} (PCF) or through the more popular \textit{distributed coordination function} (DCF). 
 
%SDN can also be used coupled with a centralized system-wide-optimizing controller running in concert with decentralized control on distributed agents that works with localized information and incentives. This can be a way to effectively marshall the network's resource pool. 

\vspace{1mm}
\subsubsection{When (\& When Not) To Use Resource Pooling}
 
Throughout this paper, we have provided many examples of how the economics of supporting a large number of users/ applications make the use of resource pooling attractive. This should not be construed to mean that anti-resource-pooling (sometimes called resource fragmentation) is without utility.  There is a need of both models (resource pooling/ fragmentation). As an example, in OFDM, the justification for anti-resource-pooling is the ease of signal processing on narrow frequency bands to achieve higher spectral efficiency. Anti-resource-pooling can help avoid `the tragedy of the commons'. Example of techniques that rely on resource fragmentation include time division multiplexing (TDM)/ frequency division multiplexing (FDM), orthogonal FDM (OFDM), time division duplex (TDD)/ frequency division duplex (FDD). The main advantage of resource fragmentation is improved service (often at a higher cost) while the main advantage of resource pooling is cost-efficient resource utilization due to the ability to utilize unused resource capacity.

\vspace{1mm}
\subsubsection{How to Fairly Share The Resource Pool}

An important question relating to resource pooling is to decide how to fairly divide the resources (i.e., who gets which share of the resource pie?). There are various fairness definitions that have been proposed \cite{lan2010axiomatic}. The naive view that fairness means absolute equality breaks down when we consider that different users have different preferences and utilities for different resources. It may also be preferable to use unequal allocation if all the users individually get a bigger share of the resource pie.  Commonly used fairness criteria include \textit{max-min fairness} (maximize the minimum share every one gets), \textit{max utilization} (the resources are maximally utilized), and \textit{(weighted) proportional fairness} in which aggregate of (weighted) proportional changes is non-positive for any other allocation). While devising an efficient resource allocation scheme, it is also important to employ an incentive-compatible scheme that encourages users and applications to choose the ``best'' action using insights from the field of ``mechanism design''. On the Internet, since user applications and end hosts are directly in charge of determining their resource share (according to the TCP protocol), end users can selfishly claim more than their fair share and thus cause a tussle between end hosts, the various stakeholders operating the network \cite{clark2002tussle}. An interesting new approach for a fairer, faster Internet \cite{briscoe2008fairer} is to allow everyone to use as much of the Internet resource pool as they can, but to police and push back those users/ applications that limit the freedom of others when ``too much load meets too little capacity''  \cite{jacquet2008policing}.

%The problem of how to best manage a common-resources pool, as can be expected, is not confined to computer networks only. There has been a lot of research in economics on how to best manage commons \cite{ostrom1990governing}, and researchers can benefit by studying these ideas and examine their applicability to manage technological resource pools. 

\subsection{Can Resource Pooling Mechanisms Conflict/ Fail?}

%Since resource pooling has so many different forms, it is difficult to make general statements about resource pooling. 

While it has been shown that resource pooling indeed does help in responding robustly to failures and overloads, we must also note that resource pooling implementations can be complex or even unstable. Some forms of resource pooling---such as dynamic multipath routing---can lead to instability and hysteresis (the phenomenon in which a physical effect lags behind changes in the effect causing it). The approach of resource pooling can fail if the various stakeholders adopt a myopic greedy approach---captured by the economic notion of the ``tragedy of the commons'' popularized by Hardin in 1968. Like other commons, unlicensed spectrum can also be plagued potentially by the ``tragedy of the commons'' in which uninhibited access by many users can render spectrum useless. It is also important to manage any ``tussle'' that results \cite{clark2002tussle} through the interaction of various stakeholders of the Internet (e.g., users and network), and to ensure that the various multi-layer resource pooling techniques combine harmoniously without any conflicts \cite{wischik2008resource}.

%It has been shown that resource pooling can produce phase transitions for amplified load and also obscure the approach to capacity limits \cite{kelly2014stochastic}. 

%There is disagreement if such tragedies can be a practical problem for wireless commons in unlicensed spectrum \cite{sicker2006examining}, but the potential of such a anomaly rising in community networks cannot be dismissed.  

%In a previous work focused on ``resource pooling'', it was stated that ``resource pooling is such a powerful tool that designers at every part of the network will attempt to build their own load-shifting mechanisms. A network architecture is effective overall, only if these mechanisms do not conflict with each other.'' \cite{wischik2008resource}. 

%Since resource pooling has diverse manifestations, and can be implemented at different layers,
% 
%As an example, the network may be implementing resource pooling optimizing for network costs while a P2P application, also implementing resource pooling, can conflict with the network policies due to myopic application-specific optimizations. 

\section{Conclusions}
\label{sec:conclusion}

The main contribution of this paper is that we identify ``resource pooling'' as a general unifying principle underpinning a wide variety of wireless technologies. We advance the state of the art by demonstrating the generality of the concept of resource pooling in wireless networks. Since the individual wireless technologies based on resource pooling have been developed independently, a plethora of piecemeal solutions have emerged that may unfortunately not work together harmoniously, and may even conflict. A principled unified understanding of resource pooling in wireless technologies can act as the first step towards avoid potentially conflicting wireless resource pooling mechanisms. We also highlight the important role such resource pooling techniques play in the developing world. 

%\section{TO BE MOVED SOMEWHERE}
%
% It is worth clarifying that while technology alone cannot solve the development problem \cite{toyama2011technology}---e.g., technology can at best only amplify, and not substitute, human willing to change---technology is, nonetheless, an important cog of the multi-faceted approach needed to tackle the problems of poverty.
% 
%Very low-cost internet access using KioskNet \cite{guo2007very}.
%
%Revisiting Resource Pooling: The Case for In-Network Resource Sharing \cite{psaras2014revisiting}

\bibliographystyle{ieeetr}
%\bibliography{RP4D}
{\footnotesize
\bibliography{RP4D}}

\end{document}